# Privacy in Federated Learning

Jaydip Sen[1], Hetvi Waghela[2], Sneha Rakshit[3]
Department of Data Science, Praxis Business School, Kolkata, INDIA
email: [1]jaydip.sen@acm.org, [2]waghelah@acm.org, [3]srakshit149@gmail.com

## 1. Introduction

Federated Learning (FL) introduces a new method in *machine learning* (ML) where the training process is decentralized, enabling multiple devices or servers to collaboratively build a model without sharing their individual data. This method greatly improves data privacy and security, making it especially important in our current data-centric environment where issues of data breaches and privacy are critical.

Google researchers introduced the concept of FL in 2016 to improve user privacy while still leveraging the advantages of large-scale ML models. The initial application was in the context of mobile devices, particularly to improve the performance of predictive text input on smartphones without compromising user privacy. The introduction of FL marked a significant shift towards privacy-preserving ML techniques and has since been adopted and refined across various industries and applications.

FL is becoming more pertinent amid stringent data privacy regulations and heightened public concern over data security. Legislations like the General Data Protection Regulation (GDPR) in Europe and the California Consumer Privacy Act (CCPA) in the USA establish rigorous guidelines governing the collection, storage, and processing of personal data. These regulations challenge the feasibility of traditional centralized data processing methods, driving the adoption of privacy-preserving techniques like FL.

Moreover, the widespread use of Internet of Things (IoT) devices, mobile phones, and edge computing has led to a massive generation of data at the network's edge. FL takes advantage of this distributed data by allowing devices to collaboratively train models locally. This approach minimizes the need to transfer large amounts of data to central servers, thereby reducing the risks associated with data breaches.

**Core Concepts and Mechanisms:** FL involves several core concepts and mechanisms that differentiate it from traditional ML approaches:

1. Local Training: Each client device independently trains a local model using its own data. This training process can be adapted to fit the device's capabilities and the nature of the data.
2. Model Updates: After training, each client computes model updates, which are essentially the changes to improve the model based on the local data. These updates are sent to a central server.
3. Aggregation: The central server collects model updates from multiple clients to enhance the global model. This aggregation can be achieved through various methods, such as weighted averaging, ensuring that the global model incorporates the collective learning from all clients.
4. Communication Protocols: Efficient communication protocols are essential in FL to minimize the overhead and latency associated with transmitting model updates between clients and the central server. Techniques like secure aggregation and *differential privacy* (DP) can be employed to enhance the security and privacy of the updates during transmission.

**Types of FL:** FL can be categorized into different types based on the nature of the clients and the data they hold:

1. Cross-Device FL: This type involves a large number of relatively lightweight devices, such as smartphones and IoT devices, each with a small amount of data. The primary challenge in cross-device FL is managing the communication and computation constraints of these devices.
2. Cross-Silo FL: This type involves a smaller number of organizations or institutions (silos) that have substantial computational resources and larger datasets. Examples include hospitals collaborating on medical research or banks working together to improve fraud detection systems. Cross-silo FL typically deals with fewer clients but larger and more heterogeneous datasets.

**Privacy-Preserving Techniques:** FL enhances privacy by keeping data local, but additional techniques can further strengthen privacy guarantees:

1. Differential Privacy (DP): By incorporating noise into the model updates, DP ensures that sensitive information about individual data points is not disclosed. This technique provides mathematical guarantees about the privacy of the data.
2. Homomorphic Encryption (HE): This technique facilitates computations directly on encrypted data, ensuring the data remains confidential throughout the entire process [1]. HE is computationally intensive but offers strong privacy protection.
3. SMPC(SMPC): It facilitates collaborative computation of a function by multiple parties using their inputs, ensuring their confidentiality. In FL, SMPC can securely aggregate model updates without exposing individual updates to the central server.
4. Secure Aggregation: This technique combines model updates such that the central server cannot view individual updates but can still calculate the overall aggregated update [2]. Secure aggregation protocols (SAPs) are crafted to safeguard the privacy of model updates during both transmission and aggregation.

**Real-World Applications:** FL has been successfully implemented in various domains, demonstrating its potential to enhance privacy while enabling collaborative ML.

1. Healthcare: In healthcare, FL enables collaboration among several hospitals to train models using patient data without disclosing them. This approach can improve diagnostic models, personalized treatment plans, and predictive analytics while complying with strict privacy regulations.
2. Finance: Financial institutions can use FL to collaboratively develop fraud detection systems, credit scoring models, and personalized financial services. By keeping customer data within each institution, FL helps maintain compliance with financial privacy regulations.
3. Mobile and Edge Devices: FL is widely used in mobile applications, such as predictive text input, personalized recommendations, and voice recognition. For instance, Google's Gboard keyboard uses FL to enhance its predictive text suggestions without transmitting user typing data.

4. Industrial IoT: In industrial IoT, FL can be applied to predictive maintenance, quality control, and supply chain optimization. Devices and sensors in different locations can collaboratively train models to predict equipment failures or optimize production processes without sharing sensitive operational data.

**Challenges in FL:** Although FL provides substantial benefits, it also poses several challenges that must be addressed. Some of them are mentioned below.

1. Data Variability: Client data can vary significantly, creating challenges in training a global model that performs well for all clients. Techniques for handling non-IID (non-independent and identically distributed) data are crucial for the success of FL.
2. Communication Overhead: Regularly communicating the updates of model between the server and clients may lead to considerable overhead, particularly in cross-device FL. Efficient communication protocols and methods to minimize the frequency and size of updates are crucial.
3. Model and Data Privacy Risks: Despite the privacy-preserving nature of FL, there are still risks of data leakage through model updates. Adversarial attacks, such as model inversion attacks, can potentially reconstruct sensitive information from model updates. Robust defense mechanisms are needed to mitigate these risks.
4. Scalability: FL needs to scale to handle millions of devices in cross-device scenarios or large datasets in cross-silo scenarios. Scalable algorithms and infrastructure are necessary to manage the complexity and scale of FL systems.

Hence, current and future research in FL is likely to focus on improving privacy guarantees, enhancing communication efficiency, developing robust defense mechanisms against adversarial attacks, and ensuring the scalability of FL systems. Advances in these areas will help realize the full potential of FL as a paradigm of ML that prioritizes privacy preservation.

FL marks a notable advancement in ML by tackling the crucial challenge of data privacy. By enabling decentralized model training, FL allows multiple entities to collaborate on improving ML models without disclosing their data. This scheme not only enhances privacy and security but also opens new possibilities for applications in healthcare, finance, mobile and edge devices, and industrial IoT. As the field of FL continues to evolve, it has the potential to transform the way we approach ML in a privacy-conscious world. With ongoing research and development, FL is poised to become a cornerstone of privacy-preserving ML, striking a balance between leveraging data-driven insights and the necessity to protect data privacy.

The organization of the chapter is as follows. Section 2 present some fundamental background information of FL. Section 3 discusses different approaches to privacy preservation of data in FL. Some of the existing approaches and schemes proposed in the literature for protecting data in FLs are presented in Section 4. Section 5 discusses some important real-world applications of FL in the healthcare, financial and electronic and embedded devices sectors, and how the privacy of critical and sensitive information are protected. Section 6 concludes the chapter highlighting some potential future work in the privacy in FL.

## 2. Fundamentals of FL

FL has evolved as a groundbreaking approach to decentralized ML, addressing the critical issue of data privacy. This section delves into the fundamentals of FL, exploring its architecture and workflow, key components, and the various types of FL.

### 2.1 Architecture and Workflow

At its core, the FL architecture involves multiple clients (e.g., mobile devices, IoT devices, or institutional servers) that collaboratively train a shared global model under the coordination of a central server. The primary innovation in FL is that the training data remains localized on the client devices, significantly mitigating privacy risks associated with traditional centralized ML. The architecture of a typical FL is depicted in Figure 1. The roles of the central server and the local clients and the work flow in FL are discussed briefly in the following.

**Central Server:** The central server has the following roles.

- *Coordination:* The central server orchestrates the overall training process, ensuring synchronization among the clients.
- *Model Initialization:* It initializes the global model parameters and disseminates them to the clients.
- *Aggregation:* The central server aggregates the model updates (gradients or parameter updates) received from the clients to form an improved global model.
- *Communication:* It handles the bidirectional communication between itself and the clients, managing the distribution of the global model and the collection of local updates.

**Local Clients:** The local clients perform the following tasks.
- *Local Data Storage:* Each client retains its data locally, ensuring that sensitive information is not exposed.
- *Local Model Training:* Clients perform training on their local data using the global model parameters received from the server.
- *Model Update Transmission:* After local training, clients compute model updates (e.g., gradients) and send them to the central server.
- *Device Heterogeneity Management:* Clients manage their computational resources to participate in the FL process, dealing with varying device capabilities and network conditions.

**Work Flow:** The FL workflow involves several iterative steps as follows.

- *Model Initialization:* The central server initializes the global model parameters and broadcasts them to all participating clients.
- *Local Training:* Clients receive the global model and train it on their local datasets. This involves forward and backward passes to compute gradients or updates specific to the client's data.
- *Model Update Transmission:* After training, each client sends its computed updates to the central server. These updates typically consist of gradient information or parameter changes derived from the local training process.
- *Global Aggregation:* The central server aggregates the received updates to form a new set of global model parameters. Common aggregation methods include averaging the updates or using more sophisticated techniques like weighted averaging, considering the size of the local datasets.
- *Model Update Broadcast:* The server disseminates the updated global model parameters back to the clients, and the process repeats for several rounds until the model converges.
- *Termination:* The FL process concludes when the global model achieves satisfactory performance metrics, such as accuracy or loss, across the participating clients.

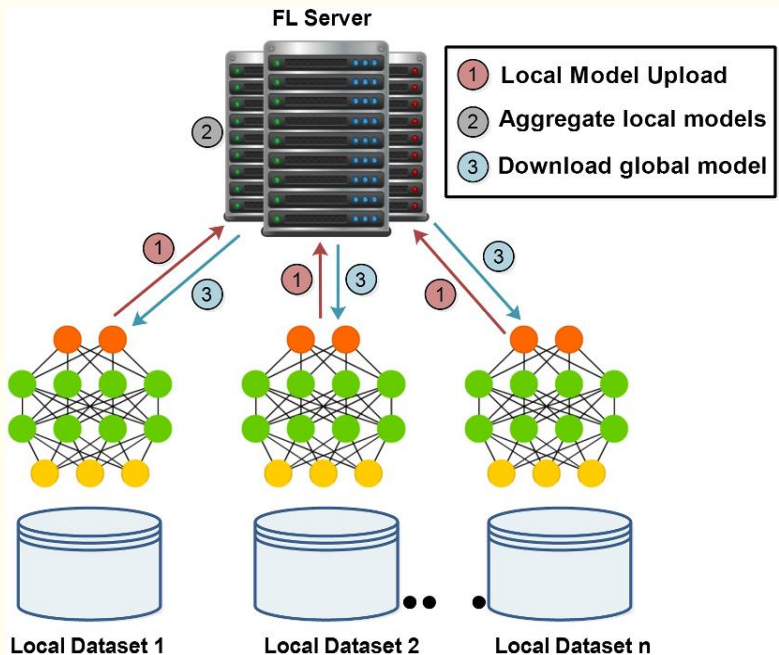

**Figure 1.** Federated learning architecture (Note: The figure is adapted from [45])

The effectiveness of FL hinges on several critical components that ensure the collaborative training process is efficient, secure, and scalable. The following components are critical for an efficient FL system.

1. **Local Training:** This involves two important components: (a) data partitioning, and (b) training algorithm. Clients utilize their local datasets, which may be non-IID (non-independent and identically distributed), meaning the data distribution varies across clients. Moreover, clients employ standard training algorithms, such as stochastic gradient descent (SGD), on their local data. The training process involves multiple epochs to minimize the local loss function.

2. **Model Update Transmission:** This involves the following tasks: (a) gradient computation, (b) update compression, and (c) secure communication. After local training, clients compute gradients representing the adjustments needed to improve the model based on their local data. To reduce communication overhead, updates can be compressed using techniques like quantization or sparsification before transmission. Secure transmission protocols ensure the confidentiality and integrity of the updates as they are sent to the central server.

3. **Global Model Aggregation:** This involves the following tasks: (a) averaging, (b) weighted averaging, and (c) advanced aggregation. The simplest and most common aggregation method is averaging the model updates from all clients. This approach assumes that each client's contribution is equally valuable. To account for varying data sizes and qualities, the server may use weighted averaging, giving more weight to updates from clients with larger or more representative datasets. More sophisticated methods, such as federated optimization algorithms, can be employed to improve convergence rates and model performance.

## 2.2 Types of FL

FL can be categorized based on the nature and scale of the clients involved. The two primary types are cross—device FL and cross-silo FL. These are discussed in the following.

### 2.2.1 Cross-Device FL

Cross-device FL involves a vast number of relatively lightweight devices, such as smartphones, tablets, and IoT devices. Each device typically has a small amount of local data and limited computational resources. This type of FL has the following characteristics: (i) massive scale, (ii) device heterogeneity, (iii) intermittent availability, and (iv) privacy sensitivity. Potentially millions of devices can participate in the training process. Devices vary widely in terms of computational power, storage capacity, and network connectivity. Devices may frequently join or leave the training process due to power constraints, network availability, and user behavior. User data on these devices often includes highly sensitive information, necessitating robust privacy-preserving mechanisms. Cross-device FL applications are typically found in mobile apps and IoT systems. In mobile apps it is mostly used for enhancing predictive test input, voice recognition, and personalized recommendations without compromising user privacy. On the other hand, for IoT systems, it finds applications in wearables, home and industrial IoT systems through collaborative learning while maintaining data confidentiality.

### 2.2.2 Cross-Silo FL

Cross-silo FL involves a smaller number of clients, typically institutions or organizations, each with substantial computational resources and large datasets. Cross-silo FL has the following characteristics: (i) limited number of clients, (ii) data homogeneity, (iii) stable participation, (iv) regulatory compliance requirements. Cross-silo FL usually involves tens to hundreds of clients. Data may be more homogeneous within each silo but can vary significantly between different silos. Since institutions have more stable and reliable participation compared to individual devices, in cross-silo FL, entities have more stable participation. However, ensuring compliance with data protection regulations, such as GDPR and Health Insurance Portability and Accountability Act (HIPAA), is critical. Cross-silo FL finds applications in healthcare, finance, and research sectors. In the healthcare sector, cross-silo FL is used in collaborative training of diagnostic models across multiple hospitals without sharing patient data. Developing fraud detection systems by leveraging data from different banks while maintaining customer privacy is a typical application of cross-silo FL in finance. In the field of research collaborations, universities and research institutions can jointly train models on sensitive data, such as genomic information, without data exposure.

## 2.3 Exploration of the Key Components of FL

This sub-section provides more details on the three key components of a federated leaning system.

*(1) Local Training:* Local training is the cornerstone of FL, as it enables each client to leverage its own data to improve the global model. The following elements are crucial to this process: (i) data partitioning, and (ii) training algorithms. In FL, data are inherently partitioned across clients. This partitioning can be either horizontal or vertical. In horizontal FL, each client has data with the same feature space but different samples (e.g., different users). In vertical FL, each client has data with different feature spaces but potentially the same samples (e.g., different institutions sharing user data with non-overlapping features). Clients use standard ML algorithms adapted to the local context. Common algorithms include: (1) stochastic gradient descent (SGD), (2) federated averaging (FedAvg), (3) personalized FL. SGD is the widely used algorithm due to its simplicity and effectiveness in large-scale optimization. FedAvg is a specific adaptation of SGB for FL, where local models are trained for multiple epochs before averaging. Personalized FL techniques allow each client to tailor the global model to better fit its local data, enhancing performance in heterogeneous environments.

*(2) Model Update Transmission:* Efficient and secure transmission of model updates is critical for the success of FL. Model update transmission involves the following issues: (1) gradient computation, (2) update compression, and (3) secure communication. Clients compute the gradients or parameter updates based on their local training. These updates encapsulate the information needed to improve the global model without exposing the raw data. To minimize communication overhead, updates can be compressed. Compression techniques include methods such as quantization, sparsification, and pruning. Quantization, however, reduces the precision of the updates as it uses fewer bits to represent each parameter. Sparsification involves sending only the most significant updates and zeroing out small updates. Pruning removes redundant parameters from the updates. Finally, ensuring the confidentiality and integrity of the updates during transmission is paramount. Encryption and secure aggregation are two methods used to preserve the confidentiality and integrity of updates. Encryption involves the use of cryptographic techniques to protect the updates in transit. SAPs are used for aggregating updates in a way that prevents the server from having access to individual contributions from the clients.

*(3) Global Model Aggregation:* The central server's role in aggregating the updates from multiple clients is vital to FL. Following approaches are generally used in global model aggregation: (1) averaging, (2) weighted averaging, and (3) advanced aggregation techniques. Averaging is the most straightforward method, where the server computes the average of all received updates. This approach assumes equal importance for all updates. Weighted averaging accounts for the size of the local datasets by giving more weight to updates from clients with larger datasets. This method helps balance the influence of different clients based on their data volume. Advanced aggregation techniques employing more sophisticated approaches, such as federated optimization and adaptive aggregation can improve the efficiency and effectiveness of aggregation. Federated optimization algorithms like Federated SGD and Federated ADAM can enhance convergence rates and model performance. Adaptive aggregation methods that adjust the aggregation method based on the characteristics of the updates received, such as considering the variance in the variance in the updates are found to be more accurate.

### 2.4 Challenges in FL

As FL continues to evolve, several areas require further research and development to address existing challenges and enhance its capabilities. Some of the critical challenges are as follows.

- *Enhanced privacy-preserving techniques:* Developing more robust privacy-preserving mechanisms, such as advanced DP techniques, HE, and secure multi-party computation, to ensure stronger privacy guarantees.
- *Improved scalability:* Creating scalable algorithms and infrastructure to handle the massive scale and diversity of devices in cross-device FL. This includes optimizing communication protocols and reducing the computational burden on resource-constrained devices.
- *Efficient model aggregation:* Innovating aggregation methods that can handle the heterogeneity of updates and improve the convergence rates of global models. Techniques like federated optimization and adaptive aggregation can play a significant role.
- *Personalized FL:* Developing methods that allow the global model to be personalized for individual clients, improving performance in heterogeneous environments. Approaches like federated meta-learning and multi-task learning can be explored.
- *Robustness and security:* Enhancing the robustness of FL systems against adversarial attacks and ensuring the security of model updates. Techniques like adversarial training and SAPs are critical.
- *Regulatory compliance:* Ensuring that FL frameworks adhere to data protection regulations across different regions. This involves continuous monitoring and updating of compliance strategies as regulations evolve.
- *Interdisciplinary collaboration:* Encouraging collaboration between researchers from different fields, such as ML, cryptography, and data privacy, to develop innovative solutions for FL.

FL represents a transformative approach to ML, addressing the critical issue of data privacy while enabling collaborative model training across distributed clients. By keeping data localized and leveraging privacy-preserving techniques, FL offers significant advantages over traditional centralized models. The architecture and workflow of FL, involving local training, model update transmission, and global model aggregation, provide a robust framework for decentralized learning.

The distinction between cross-device and cross-silo FL highlights the versatility of FL in different contexts, from personal devices to institutional collaborations. Each type of FL presents unique challenges and applications, necessitating tailored solutions to optimize performance and privacy.

As FL continues to advance, ongoing research and development in privacy-preserving techniques, scalability, model aggregation, and regulatory compliance will be crucial to realizing its full potential. By fostering interdisciplinary collaboration and addressing existing challenges, FL can pave the way for a privacy-centric approach to ML that empowers individuals and organizations while driving innovation and collaboration.

### 3. Privacy-Preserving Techniques in FL

Privacy-preserving techniques in FL are crucial for protecting the confidentiality of data while enabling collaborative model training. This section delves into several key methods, including DP, encryption methods, secure aggregation, and anonymization/pseudonymization, to ensure privacy and security in FL systems. In this section, these methods are discussed briefly.

## 3.1 Differential Privacy

Differential privacy (DP) is a formal privacy framework designed to provide strong guarantees that individual data points in a dataset cannot be distinguished from each other. This is achieved by introducing randomness into the data or computations. The goal is to ensure that the output of a computation (e.g., a model update) does not reveal whether any single individual's data was included in the input, thereby preserving privacy.

In the context of FL, DP can be applied to the model updates sent from the clients to the central server. This typically involves adding noise to the updates to obscure the contributions of individual data points. Depending on the type of noise being added and the way the noise is added, different types of DP may be implemented as discussed in the following.

***Noise addition:*** Each client adds noise to its computed gradients or parameter updates before sending them to the central server. The noise is typically drawn from a statistical distribution, such as a Gaussian or Laplace distribution, with a scale determined by a privacy parameter (epsilon, ε). A smaller ε indicates stronger privacy guarantees but may reduce the utility of the model.

***Local Differential Privacy:*** This approach ensures that each client's data remains private even before aggregation. The noise added at the client level is calibrated to provide DP guarantees. This method protects against adversaries who might intercept updates during transmission.

***Global Differential Privacy:*** In some cases, noise is added at the server level after aggregating the updates from all clients. This ensures that the aggregated update meets DP guarantees, though it may require trusting the server to some extent.

***Advantages and Challenges in DP:*** DP provides two advantages: (i) strong privacy guarantees and (ii) flexibility. DP provides mathematically proven rigorous privacy guarantees and it can be adjusted to trade-off between privacy and model accuracy. However, there are some challenges which include: (i) accuracy vs privacy trade-off and (ii) hyperparameter tuning. Adding noise can degrade the model's performance, especially if the privacy requirements are stringent. Selecting appropriate noise scales and privacy parameters requires careful tuning and domain knowledge.

## 3.2 Encryption Methods

Encryption methods are essential for ensuring that data and model updates remain confidential during transmission and computation. Two prominent techniques in FL are *HE* and *secure multi-party computation*. These techniques are discussed briefly in the following.

***(1) Homomorphic Encryption:*** HE is a type of encryption that allows computations to be performed on ciphertexts (encrypted data) without needing to decrypt them. The result of the computation remains encrypted and can be decrypted later to obtain the correct result. HE in FL involves three fundamental steps; (i) encryption of updates, (ii) aggregation of encrypted updates, and (iii) decryption. Clients encrypt their model updates using a HE scheme before sending them to the central server. Common schemes include partially HE (PHE) and fully HE (FHE). The central server aggregates the encrypted updates directly, performing operations like addition and multiplication on the ciphertexts. Since the operations are homomorphic, the result is an encrypted aggregate that can be decrypted by an authorized party. After aggregation, the server or a trusted third party decrypts the aggregated results to obtain the updated global model.

***Advantages and Challenges in DP:*** Confidentiality and security are the two distinct advantages of HE. HE ensures that updates remain confidential throughout the computation process. It also protects against external adversaries and malicious servers. However, there are challenges associated with HE too. Two important challenges are: (i) high computational overhead and (ii) high complexity. HE schemes, particularly FHE, can be computationally intensive and slow. Moreover, implementing HE requires significant expertise and careful handling of cryptographic parameters.

***(2) Secure Multiparty Computation (SMPC):*** SMPC allows multiple parties to jointly compute a function over their inputs while keeping those inputs private. The computation is structured so that no party learns anything about the other parties' inputs beyond what can be inferred from the output. SMPC involves three steps in its operation in FL. These steps are: (i) secret sharing, (ii) distributed computation, and (iii)reconstruction. Each client splits its model updates into multiple shares and distributes them among the participating parties (including the central server and other clients). The parties collectively perform the aggregation on the shares. No single party has enough information to reconstruct the original updates during the computation. After computation, the shares are combined to obtain the aggregated update, which is then used to update the global model.

***Advantages and Challenges in SMPC:*** The primary advantages of SMPC are: (i) high level of privacy and (ii) no need of trusted environment. SMPC ensures that no individual party learns the complete updates from any other party. It also reduces the need to trust any single party, enhancing overall security. However, SMPC has its own challenges too. SMPC typically requires significant communication between parties, which can be a bottleneck. Moreover, designing and implementing SMPC protocols is complex and requires careful coordination.

### 3.3 Secure Aggregation

Secure aggregation is a technique designed to aggregate model updates in such a way that the central server cannot see individual contributions. Instead, it only sees the aggregated result, ensuring the privacy of individual updates. In FL, SAPs work in three steps: (i) encryption of updates, (ii) aggregation of encrypted updates, and (iii) decryption. In the first step, each client encrypts its model updates using a secure encryption scheme before sending them to the central server. In the second step, the central server aggregates the encrypted updates. The protocol ensures that the server can only decrypt and aggregated result and not the individual updates. In the third and final step, the aggregated update is decrypted, providing the global model update without revealing individual contributions.

The implementation of SAPs involves two initial steps: (i) pairwise masking and (ii) additive secret sharing. In the pairwise masking phase, each client generates a random mask and shares it with other clients using a secure channel. The masks are used to obfuscate the updates before sending them to the server. The server aggregates the masked updates, and the masks cancel out in the aggregation process, revealing the aggregated results. In the additive secret sharing phase, each update is split into multiple shares, and the shares are distributed among multiple servers or parties. The servers perform the aggregation on the shares, ensuring that no single server learns the individual updates.

***Advantages and Challenges in Secure Aggregation:*** Secure aggregation has two distinct advantages: (i) enhanced privacy and (ii) higher efficiency. Secure aggregation ensures that the server cannot see individual updates, enhancing privacy. Moreover, compared to full HE or SMPC, secure aggregation can be more efficient in terms of computation and communication overheads. Higher complexity and a lack of robustness are two challenges for secure aggregation. Implementing SAPs requires careful design and coordination among clients. Additionally, the protocol must handle cases where some clients drops out of behave maliciously.

### 3.4 Anonymization and Pseudonymization

Anonymization and pseudonymization are techniques used to obscure personal identifiers in data, making it difficult to link data back to specific individuals. While these techniques are commonly used in data privacy, they also play a role in FL to enhance the privacy of participants.

***Anonymization:*** Anonymization encompasses various techniques including data anonymization, *K*-anonymity, and *L*-diversity. Data anonymization involves removing or altering personal identifiers, such as names, addresses, and other unique attributes, to prevent the data from being traced back to individuals. *K*-anonymity ensures that each record in the dataset is indistinguishable from at least $K$-1 other records based on certain identifying attributes, thereby minimizing the risk of re-identification. *L*-diversity builds on *K*-anonymity by ensuring that each group of similar records (equivalence class) contains at least *L* different values for sensitive attributes, offering enhanced protection against attribute disclosure.

***Pseudonymization:*** In data pseudonymization, identifiers are replaced with pseudonyms or tokens that can only be linked back to the original identifiers using a separate mapping table. This ensure that data analysis can be performed without revealing personal identifiers. Unlike anonymization, pseudonymization is reversible, allowing data to be reidentified, if necessary, using the mapping table.

***Advantages and Challenges in Anonymization and Pseudonymization:*** There are two advantages for these two approaches: (i) enhanced privacy and (ii) compliance with standards. Anonymization and pseudonymization protect personal identifiers. Moreover, these techniques help in meeting regulatory requirements for data protection, such as GDPR and HIPAA. However, there are challenges too. Anonymization can lead to loss of data utility, making it harder to perform certain analyses. Again, pseudonymization is reversible, which means it requires secure handling of the mapping table to prevent data breaches.

Privacy-preserving techniques in FL are essential for ensuring the confidentiality and security of data while enabling collaborative model training. DP, encryption methods like HE and secure multi-party computation, secure aggregation, and anonymization/pseudonymization each plays a crucial role in protecting privacy.

DP provides strong mathematical guarantees by adding noise to model updates, ensuring that individual data points remain indistinguishable. Encryption methods like HE allow computations on encrypted data, while SMPC enables collaborative computation without data leakage. SAPs ensure that the server can only see the aggregated result, not individual updates. Anonymization and pseudonymization techniques obscure personal identifiers, further enhancing privacy and compliance with data protection regulations.

Each technique has its own benefits and challenges, and their implementation involves balancing trade-offs between privacy, utility, and computational overhead. As FL continues to evolve, ongoing research and development in these privacy-preserving techniques will be crucial to addressing existing challenges and enhancing the security and effectiveness of FL systems.

## 4. Existing Methods of Privacy in FL

Privacy in FL (FL) has become a significant area of research due to the sensitive nature of data involved and the increasing concern over data privacy. This section surveys important existing works that have addressed privacy issues in FL, encompassing various techniques and methodologies. We will explore DP, encryption methods, SAPs, and other privacy-preserving mechanisms that have been proposed or implemented in FL.

### 4.1 Differential Privacy in FL

In this section, some DP-based privacy scheme for FL are discussed in details.

McMahan et al. introduce a method for training recurrent language models using DP within the FL framework [3]. The authors propose using differentially private stochastic gradient descent (DP-SGD) to ensure that individual user data remains protected and cannot be reverse-engineered from the trained model. By introducing noise adding noise into model updates, this method provides DP guarantees while trading-off privacy with the accuracy of model. Demonstrated on a language modeling task, the technique predicts the next word in a sequence based on previous words, enabling collaborative learning from a large user base while maintaining text data privacy.

Abadi et al. presents a framework for training deep learning models with strong privacy guarantees using DP [4]. The authors introduce *differentially private stochastic gradient descent* (DP-SGD), which adds carefully calibrated noise to the gradients during the training process to ensure that individual data points cannot be identified. The scheme introduces privacy accounting techniques that track the cumulative privacy loss over multiple training iterations, known as the privacy budget. Empirical results demonstrate the effectiveness of DP-SGD on various computationally intensive tasks based on deep learning.

Geyer et al. propose a novel scheme integrating DP into FL at the client level [5]. The authors propose adding noise to the updates from each client before they are aggregated, ensuring that the central server cannot infer information about any individual client's data. The authors also provide a comprehensive analysis of the privacy budget and its implications on model performance.

Fu et al. provide an analysis of the integration of DP in FL to improve security and privacy of data [6]. The authors systematically review existing methodologies that combine DP techniques with FL frameworks, addressing the inherent privacy risks. They also discuss how DP ensures that the inclusion or exclusion of any single participant's data does not significantly alter the results protecting data privacy. The work categorizes various approaches based on their implementation strategies, such as noise addition, gradient perturbation, and secure aggregation. The authors also provide a critical evaluation of the scalability and efficiency of these methods, considering the computational and communication overheads involved.

Gu et al. investigate how the integration of DP affects the fairness of ML models in FL settings [7]. The authors explore the tension between ensuring privacy and maintaining model fairness, highlighting that privacy-preserving techniques can inadvertently introduce biases. They systematically analyze the impact of DP on model performance across different demographic groups, identifying potential disparities. Their findings suggest that while DP effectively protects individual data, it can also exacerbate inequalities in model predictions, leading to unfair outcomes for certain groups. The work also discusses various metrics for assessing fairness and evaluates the trade-offs involved in balancing privacy and fairness. The authors propose methods to mitigate the adverse effects on fairness, such as adaptive noise mechanisms and fairness-aware training algorithms.

Li et al. present an advanced framework for improving the privacy of FL systems [8]. The proposition includes a novel optimization scheme that integrates DP to protect individual user data during the collaborative training process. By introducing an adaptive noise mechanism, the framework dynamically adjusts the noise added to the updates, to optimally trade-off privacy and accuracy of models. This approach mitigates the performance degradation typically associated with DP. The work also introduces a SAP to ensure that only the aggregated results are accessible, further safeguarding individual contributions. Experimental evaluations on various datasets demonstrate that their optimized scheme significantly improves model performance while maintaining strong privacy protection.

Löbner et al. explore the application of *local DP* (LDP) to protect user data in FL scenarios, specifically for email classification tasks [9]. A new scheme is proposed that integrates LDP into the FL process, ensuring that users' raw data remains private even before it is transmitted for aggregation. By applying noise to the data at the local level, their method prevents sensitive information from being exposed during model training. The framework effectively addresses privacy concerns inherent in FL, in which a model is trained in a collaboratively way using data from multiple sources. The work presents a detailed analysis of the trade-offs between privacy and model accuracy, demonstrating that their approach maintains high classification performance while providing robust privacy guarantees. Experimental results on email datasets illustrate that the LDP-enhanced FL model can achieve competitive accuracy compared to traditional methods.

Wei et al. delve into the development and evaluation of DP-enhanced algorithms for FL [10]. The authors propose a suite of algorithms that incorporate DP mechanisms to safeguard individual data contributions during the FL process. The trading-off of privacy and accuracy of models has been done ensuring that the utility of the trained models remains high while providing strong privacy guarantees. The work details the mathematical foundations of the proposed algorithms, including the specific noise addition techniques used to achieve DP. Through comprehensive theoretical analysis, the authors establish the privacy guarantees and performance bounds of their algorithms. They also present extensive empirical evaluations on various benchmark datasets, demonstrating that their methods maintain competitive accuracy compared to non-private FL approaches. The results highlight the effectiveness of their algorithms in mitigating privacy risks without significantly degrading model performance.

Li et al. introduce a novel approach that combines FL with transfer learning while incorporating DP to protect sensitive data [11]. The authors aim to address the challenge of training models collaboratively across different organizations that have diverse datasets, without compromising privacy. Their framework leverages transfer learning to enable knowledge transfer from a source domain to a target domain within a FL setup. To ensure privacy, they integrate DP mechanisms, adding noise to the model updates to prevent the exposure of individual data points. This combination allows organizations to benefit from shared knowledge without the need to share raw data, preserving both privacy and data utility. The work also provides a theoretical analysis of the privacy guarantees and evaluates the performance of the proposed method through experiments on real-world datasets. The results demonstrate that their approach maintains high model accuracy while providing strong privacy protection.

Park & Choi explore the integration of DP in FL systems that utilize *over-the-air computation* (OAC) [12]. The scheme exploits the inherent properties of OAC to enhance privacy and efficiency in FL. By combining OAC with DP, the framework ensures that individual data contributions remain confidential during the aggregation process. The approach uses OAC to simultaneously aggregate updates from multiple devices over a wireless channel, adding noise to the aggregated signal to achieve DP.

This scheme has a reduced overhead of computing and communication making it scalable for large-scale FL deployments.

## 4.2 Encryption Methods in FL

Encryption methods play a vital role in securing data during the FL process. This section discusses some encryption-based schemes for FL privacy.

Keith Bonawitz et al. introduces a protocol designed to enhance privacy in FL by securely aggregating user-held data [13]. The authors address the challenge of ensuring that individual users' data remains confidential while still enabling the collective training of a ML model. Their approach uses cryptographic techniques to perform secure aggregation, ensuring that only the aggregated results are revealed, not the individual contributions. This is achieved through a combination of HE and secret sharing, which allows the aggregation process to be both secure and efficient. The protocol is robust against dropouts, meaning it can handle the scenario where some users do not complete the training process. Furthermore, it is designed to be scalable, accommodating many participants with minimal overhead.

Truex et al. explore the integration of several privacy-preserving schemes for FL to enhance the security and efficiency [14]. The authors recognize that no single approach is sufficient to address all privacy and scalability challenges, thus advocating for hybrid solutions. They combine DP, secure multiparty computation, and HE to protect sensitive data during the FL process. Secure multiparty computation enables multiple parties to collaboratively compute a function value based on their individual inputs which are private to them. HE allows computations to be carried out on encrypted data without needing decryption. The work also discusses the several optimization techniques for trade-offs computing and communication overhead with the level of privacy achieved.

Phong et al. revisit existing methods and propose enhancements to strengthen the privacy of deep learning models [15]. The authors address the challenge of protecting sensitive data during the training process by leveraging advanced cryptographic techniques. They build upon HE to allow computations on encrypted data, ensuring that data privacy is maintained without exposing underlying information. The proposed enhancements focus on optimizing the encryption schemes to mitigate the significant computational overhead typically associated with HE. By doing so, they make privacy-preserving deep learning more practical for real-world applications. The paper also introduces methods to maintain model accuracy while ensuring privacy, balancing the trade-offs between privacy protection, accuracy, and computational efficiency.

Park & Lim explore the implementation of privacy-preserving FL (FL) using HE [16]. The authors propose a method ensures that sensitive information remains secure during the training process. They also address the challenges associated with integrating HE into FL, such as computational overhead and communication costs. To tackle these, the authors propose optimizations that balance privacy, efficiency, and accuracy. Detailed experimental results demonstrating the feasibility and effectiveness of the proposed approach are also presented.

Kurniawan & Mambo investigates the use of HE to enhance privacy preservation in FL, specifically for deep active learning (DAL) scenarios [17]. The proposed technique ensures data privacy is protected during model training. The authors identify several challenges of applying HE in the context of deep active learning, such as increased computational demands and communication overhead. The work also proposes several optimizations to mitigate these challenges, balancing security with efficiency and performance. Experimental results validate the feasibility and effectiveness of their approach, demonstrating that it can maintain high levels of data privacy without significantly compromising the learning outcomes. The authors' method provides a practical solution for secure collaborative learning, particularly in environments where data sensitivity is a primary concern.

Nguyen & Thai addresses the critical issue of preserving privacy and security in FL [18]. The authors examine various privacy and security threats inherent in FL, such as data leakage, model inversion attacks, and malicious participants. They propose a comprehensive framework that incorporates multiple techniques to mitigate these risks, including DP, secure multi-party computation, and robust aggregation methods. The framework proposed by the authors aims to protect both the data and the model parameters during the training process. Experimental evaluations demonstrate the effectiveness of the proposed framework in maintaining privacy and security without significantly degrading model performance.

Gao et al. explore strategies for ensuring privacy and reliability in decentralized FL [19]. The authors address critical issues related to privacy preservation and reliability in FL environments. They propose a novel framework that integrates privacy-preserving techniques such as DP and secure multiparty computation to safeguard sensitive data during the learning process. Additionally, the framework incorporates mechanisms to enhance reliability, ensuring the robustness of the FL system against potential failures and malicious attacks. The proposed methods are designed to protect both the data and model integrity, thereby enhancing the overall security of the system. Experimental results validate the effectiveness of the framework, demonstrating that it can maintain high levels of privacy and reliability without compromising the performance of the learning model.

Mothukuri et al. present a comprehensive survey on the security and privacy challenges in FL [20]. The authors systematically review the various security and privacy threats that can affect FL, such as data poisoning, backdoor attacks, and inference attacks. They discuss existing defense mechanisms, including DP, secure multiparty computation, and HE, highlighting their strengths and limitations. The survey also explores the balance between model performance and the robustness of these security measures. The authors emphasize the importance of designing scalable and efficient solutions to address the evolving threats in FL environments. They identify gaps in the current research and suggest potential directions for future work to enhance the security and privacy of FL.

Zhao et al. address the challenge of maintaining efficiency and privacy in FL while defending against poisoning adversaries [21]. The decentralized nature of FL makes it vulnerable to poisoning attacks, where malicious participants can corrupt the model by injecting false data. The authors propose a robust framework that combines privacy-preserving techniques with mechanisms to detect and mitigate poisoning attacks. Their approach employs DP to protect individual data contributions and integrates anomaly detection algorithms to identify and exclude malicious updates. Experimental evaluations demonstrate the effectiveness of the proposed methods in enhancing both the security and accuracy of FL models.

Wang et al. introduce VOSA, a framework designed to enhance privacy-preserving FL through verifiable and oblivious secure aggregation [22]. FL enables collaborative model training across decentralized devices, ensuring data privacy by keeping data local. However, the aggregation of local updates poses privacy risks and requires secure methods to prevent data leakage. VOSA addresses these concerns by integrating secure aggregation techniques with verifiable computation, ensuring that the aggregated results are both accurate and privacy-preserving. The framework leverages cryptographic protocols to perform oblivious aggregation, meaning that the server cannot learn individual contributions. Additionally, VOSA includes mechanisms for participants to verify the correctness of the aggregation process, enhancing trust and reliability. Experimental results demonstrate that VOSA effectively maintains privacy and security without significantly impacting the efficiency of the FL process.

## 4.3 Secure Aggregation Protocols (SAPs)

SAPs ensure that the central server can aggregate model updates without accessing individual updates, providing a layer of security that protects user data.

Zhao et al. introduces SEAR, a novel framework designed to enhance the security and efficiency of FL in the presence of Byzantine adversaries [23]. The authors address the challenge of maintaining robust model performance when some participants may act maliciously or send incorrect data. SEAR combines secure aggregation techniques with Byzantine-robust algorithms to ensure that the aggregation process is both confidential and resilient to adversarial behavior. The framework employs cryptographic methods to protect data during transmission, ensuring that individual contributions remain private. Additionally, SEAR incorporates robust aggregation rules that can effectively identify and mitigate the impact of malicious participants. The authors provide a detailed analysis of SEAR's theoretical security guarantees and its practical performance.

So et al. present an innovative approach to overcoming the computational inefficiencies associated with secure aggregation in FL [24]. The authors introduce TURBO-AGGREGATE, a novel protocol designed to reduce the quadratic communication and computation costs that typically hinder scalable secure FL. This protocol leverages advanced cryptographic techniques to enable efficient aggregation while maintaining strong privacy guarantees for individual users' data. TURBO-AGGREGATE achieves its efficiency by using a hybrid approach that combines HE with a secure shuffling mechanism, significantly reducing the overhead compared to traditional methods. The authors provide a rigorous theoretical analysis of the protocol's security and performance, demonstrating that it can securely aggregate data with linear communication complexity.

Rathee et al. introduce ELSA, a secure aggregation framework for FL designed to withstand the presence of malicious actors [25]. FL allows multiple devices to collaboratively train a model without sharing their local data, preserving privacy. However, the aggregation process is vulnerable to attacks from malicious participants who may attempt to disrupt the learning process or infer sensitive information. ELSA addresses these issues by incorporating cryptographic techniques to securely aggregate model updates while ensuring that the contributions of individual participants remain confidential. The framework uses a combination of HE and zero-knowledge proofs to provide strong privacy guarantees and detect any malicious behavior. Experimental results demonstrate that ELSA effectively secures the aggregation process, maintaining model accuracy even in the presence of adversarial actors.

Fereidooni et al. introduces SAFELearn, a framework aimed at ensuring secure aggregation in private FL [26]. The authors argue that the aggregation of local model updates poses a significant risk of data leakage. SAFELearn addresses this by employing cryptographic techniques to securely aggregate the updates while ensuring that individual data contributions remain confidential. The framework leverages HE and secure multiparty computation to provide strong privacy guarantees. It also includes mechanisms to verify the integrity of the aggregated results, enhancing the overall security of the learning process. Experimental evaluations show that SAFELearn maintains model accuracy and efficiency while providing robust protection against data breaches.

Zhong et al. introduce WVFL, a framework for weighted verifiable secure aggregation in FL [27]. In FL, the aggregation of model updates is vulnerable to data leakage and tampering. WVFL addresses these issues by incorporating secure aggregation techniques with weighted updates to reflect the varying importance of different participants' data. The framework employs cryptographic protocols to ensure that the aggregation process is both secure and verifiable, preventing malicious actors from tampering with the results. Additionally, WVFL includes mechanisms to verify the correctness of the aggregated updates, enhancing trust and reliability. Experimental results demonstrate that WVFL effectively maintains the privacy and security of the aggregated data while preserving model accuracy and efficiency.

Zhou et al. present a comprehensive survey on security aggregation techniques, focusing on their application in various domains including FL and distributed systems [28]. Aggregation plays a critical role in combining data or computations from multiple sources while preserving confidentiality and integrity. The authors systematically review different approaches to secure aggregation, such as cryptographic methods like HE, secure multiparty computation, and zero-knowledge proofs. They discuss the strengths and limitations of each technique in ensuring data privacy and preventing attacks such as data leakage and manipulation. The survey also explores recent advancements and emerging trends in secure aggregation, highlighting their implications for improving the robustness and efficiency of distributed systems.

Sami and Güler explore the implementation of secure aggregation specifically tailored for clustered FL [29]. The aggregation of model updates in FL can be vulnerable to privacy breaches and attacks from malicious participants. The authors propose a novel framework that introduces clustering techniques to enhance both the efficiency and security of aggregation in federated settings. Their approach leverages cryptographic protocols such as HE and SMPC to ensure that model updates from clustered devices are aggregated securely without revealing individual contributions. The framework also includes mechanisms for verifying the integrity and authenticity of the aggregated results, thereby enhancing trust and reliability in the FL process. Experimental evaluations demonstrate that their method effectively balances privacy, security, and computational efficiency, making it suitable for practical deployment in clustered FL scenarios.

Liu et al. address the challenge of fast and secure aggregation in privacy-preserving FL [30]. The method aims to accelerate the aggregation process without compromising privacy. It leverages cryptographic techniques such as HE and SMPC to ensure that aggregated results remain confidential and accurate. The framework includes optimizations to reduce computational overhead, enabling efficient aggregation even with a large number of participating devices. Experimental results demonstrate that their method achieves significant improvements in aggregation speed while maintaining robust privacy guarantees.

Truong et al. present a comprehensive survey focused on privacy preservation in FL, specifically examining it through the lens of GDPR [31]. The authors systematically review the challenges and strategies related to privacy in FL emphasizing GDPR compliance as a critical consideration for data protection in European contexts. They discuss various privacy-preserving techniques employed in FL, including DP, FL-specific encryption methods, and anonymization techniques. The survey highlights the intersection of FL with GDPR principles such as data minimization, purpose limitation, and accountability, providing insights into how FL systems can align with regulatory requirements. Additionally, the authors explore emerging trends and future directions for enhancing privacy in FL systems under GDPR guidelines.

Li et al. provide a comprehensive survey on data security and privacy-preserving techniques in FL tailored for the edge and IoT environments [32]. The authors systematically review the unique challenges and existing solutions related to data security and privacy in FL at the edge and IoT levels. They discuss various security threats such as data leakage, inference attacks, and model poisoning, emphasizing the vulnerabilities inherent in edge devices with limited resources. The survey covers a range of privacy-preserving techniques applicable to FL, including DP, HE, secure aggregation, and FL-specific optimizations. Furthermore, the authors examine the integration of these techniques with edge computing paradigms to enhance both security and efficiency in FL systems.

## 4.4 Anonymization and Pseudonymization Techniques

Anonymization and pseudonymization are crucial for protecting personal identifiers in data, ensuring that sensitive information cannot be traced back to individuals.

Shokri & Shmatikov introduce a pioneering approach to training deep learning models on private data without compromising individual privacy [33]. The authors propose a novel framework that allows multiple participants to collaboratively train a neural network while ensuring that their training data remains confidential. This is achieved through a technique called *selective gradient sharing*, where participants only share a subset of their model updates, rather than their raw data, during the training process. These updates are further protected using DP, ensuring that the shared gradients do not reveal sensitive information about the individual data points. The framework effectively balances the trade-off between privacy and model utility, maintaining high model accuracy while providing strong privacy guarantees. The authors also address scalability by designing the system to efficiently handle many participants. Extensive experiments demonstrate that the proposed method can train deep learning models with a minimal loss in accuracy compared to standard training methods.

Rieke et al. explore the transformative potential of FL in the healthcare sector [34]. The authors highlight how FL enables the training of ML models on decentralized data, preserving patient privacy by keeping data localized on healthcare providers' servers. This approach mitigates the legal and ethical concerns associated with sharing sensitive health data. By collaborating on a global scale, healthcare institutions can leverage diverse datasets to improve model accuracy and generalizability, leading to better diagnostic tools and treatment plans.

Kaissis et al. focus on secure, privacy-preserving, and federated ML methods specifically applied to medical imaging [35]. Medical imaging datasets are often sensitive and subject to strict privacy regulations, making traditional centralized approaches challenging. FL offers a decentralized paradigm where models are trained across institutions without sharing raw data, thereby preserving patient privacy. The authors review the application of FL in medical imaging, emphasizing techniques such as DP, secure aggregation, and encryption methods tailored for healthcare settings. They discuss the benefits of FL in enabling collaborative model training across distributed datasets while complying with regulatory frameworks like GDPR and HIPAA.

Kanwal et al. address the challenge of balancing privacy concerns with the advancement of artificial intelligence in the context of histopathology for biomedical research and education [36]. Histopathological data is rich in information crucial for medical diagnostics and research but is inherently sensitive due to its potential to reveal patient identities. The authors focus on anonymization techniques aimed at safeguarding patient privacy while enabling meaningful analysis and AI model training. They review various anonymization methods applicable to histopathological images, such as pixelization, blurring, and generative models that synthesize realistic yet privacy-preserving images. The work also discusses the trade-offs between anonymization effectiveness and data utility, emphasizing the importance of preserving diagnostic accuracy and research value.

Choudhury et al. explore methods to enhance privacy in FL by anonymizing data [37]. The authors discuss various anonymization techniques, such as DP, which add noise to data to obscure individual contributions while maintaining overall utility. They also explore methods like data generalization and *k*-anonymity to protect identities within datasets. The work also examines the trade-offs between the degree of anonymization and the accuracy of the trained models, aiming to find a balance that maintains both privacy and model performance. Experimental results show that appropriate anonymization can significantly reduce privacy risks without severely impacting the learning outcomes.

Almashaqbeh & Ghodsi introduce AnoFel, a framework designed to support anonymity in privacy-preserving FL [38]. AnoFel addresses privacy concerns in FL by incorporating advanced anonymization techniques to enhance participant privacy without compromising the integrity and utility of the learned model. The framework employs cryptographic methods, such as HE and secure multiparty computation, to anonymize data contributions while allowing accurate aggregation. AnoFel also integrates DP to add an extra layer of protection against inference attacks. The authors present experimental results demonstrating that AnoFel effectively maintains high model accuracy while providing robust anonymity and privacy guarantees.

Zhao et al. focus on developing a framework for anonymous and privacy-preserving FL tailored to industrial big data applications [39]. The authors address the privacy risks associated with FL by proposing advanced anonymization techniques to safeguard individual data contributions. Their framework leverages DP and SMPC to ensure that data remains anonymous and protected during the aggregation process. This work highlights the unique challenges posed by industrial big data, such as the need for scalability and efficiency in handling large datasets. Experimental results demonstrate that their approach maintains high model accuracy while providing robust privacy and anonymity guarantees.

Agiollo et al. introduce a novel approach to FL called Anonymous FL via Named-Data Networking (NDN) [40]. The authors propose leveraging NDN to enhance anonymity and privacy in FL, as NDN focuses on content rather than data sources, thus naturally obfuscating the participants' identities. The proposed framework incorporates cryptographic techniques to secure data exchanges and ensure that model updates remain anonymous throughout the learning process. Experimental results demonstrate that their NDN-based approach effectively preserves privacy without compromising the efficiency and accuracy of the FL model.

Kobsa & Schreck explore the use of pseudonymity as a method for enhancing privacy in user-adaptive systems [41]. User-adaptive systems tailor their functionality and content to individual users, often requiring extensive personal data to do so. The authors argue that while such systems improve user experience, they also pose significant privacy risks. They propose pseudonymity as a solution, where users interact with the system under pseudonyms rather than their real identities. This approach allows users to benefit from personalization while minimizing the exposure of their personal information. The work also discusses various pseudonymity techniques and their effectiveness in protecting user privacy.

Gu et al. provide a comprehensive review of privacy enhancement methods for FL in healthcare systems [42]. The authors discuss the unique privacy challenges in healthcare FL, such as sensitive patient information and strict regulatory requirements like HIPAA and GDPR. They review various privacy-preserving techniques, including DP, which adds noise to data to obscure individual contributions, and HE. The work also covers secure multi-party computation, enabling multiple parties to jointly compute a function without revealing their inputs, and federated averaging algorithms designed to mitigate privacy risks.

## 5. Real-World Applications of FL

FL (FL) is rapidly gaining traction across various industries due to its ability to leverage decentralized data while preserving privacy. This section explores the real-world applications and case studies of FL in healthcare, finance, mobile and edge devices, and highlights specific implementations like Google's Gboard and collaborative healthcare research projects.

## 5.1 Healthcare Sector

Healthcare is one of the most promising fields for the application of FL due to the sensitive nature of medical data and the potential for improved patient outcomes through collaborative research and development.

### 5.1.1 Collaborative Research and Development

The healthcare sector stands to benefit significantly from FL due to the collaborative potential it offers while ensuring the privacy and security of sensitive medical data. This approach facilitates collaborative research and development among various healthcare institutions, leading to enhanced medical insights, improved diagnostic tool, and better patient outcomes.

FL allows multiple healthcare institutions to collaborate on research projects without sharing their data directly. This is particularly important in healthcare, where a patient data privacy is paramount, and regulations like the HIPAA and the GDPR impose strict controls on data sharing.

**Medical Imaging:** FL allows hospitals and medical institutions to collaboratively train models on medical imaging data (e.g., MRI, CT scans) without transferring patient data off-site. This leads to the development of more robust and accurate diagnostic tools. For example, models can be trained to detect tumors, fractures, and other anomalies more effectively by pooling data from multiple sources.

**Genomic Research:** Genomic data is highly sensitive and often subject to strict privacy regulations. FL enables researchers to build predictive models for genetic diseases and personalized medicine by aggregating insights from data distributed across different research centers and biobanks.

**Electronic Health Records (EHRs):** EHRs contain vast amounts of patient information that can be used to predict patient outcomes, optimize treatment plans, and identify potential health risks. FL facilitates the development of predictive models that can analyze EHRs from multiple hospitals without compromising patient privacy.

### 5.1.2 Maintaining Patient Privacy

Maintaining patient privacy is paramount in healthcare applications of FL due to the highly sensitive nature of medical data. FL addresses this concern by implementing several advanced techniques that ensure data privacy and security while still enabling collaborative research and model training. Here are some key approaches used to maintain patient privacy. Methods such as DP, secure multiparty computation (SMPC), HE, federated averaging, secure aggregation, and anonymization can all be useful in maintaining privacy of patient data.

SMPC and HE enable different healthcare institutions to collaboratively train a ML model without revealing their individual datasets to each other. Each participating institution encrypts its local model updates before sending them to the central server. The central server performs computations on these encrypted updates and aggregates them to improve the global model. By ensuring that raw data never leaves the local institution and remains encrypted during processing, SMPC provides a robust mechanism to protect patient privacy. DP adds random noise to the patient data or model updates from each institution before sending them to the central server . This added noise obscures individual datapoints, making it impossible to infer specific patient information from the aggregated model.

Federated Averaging (FedAvg) aggregated model updates from multiple clients (e.g., hospitals) in a privacy-preserving manner. Local models are trained on patient data within each institution. The resulting updates (model parameters) are sent to a central server, which averages these updates to form a new global model. Since only model parameters are shared and not the actual patient data, FedAvg significantly reduces the risk of data breaches and maintains patient privacy.

Anonymization removes all personal identifiable information (PII) from the data, making it impossible to link the data back to specific patients. Pseudonymization replaces personal identifiers with pseudonyms, allowing for indirect identification while still protecting patient privacy.

### 5.2  Financial Sector

In the finance sector, FL addresses the critical need to protect sensitive financial data while improving the accuracy and robustness of models used for various applications.

### 5.2.1  Improving Fraud Detection Algorithms

**Fraud Detection:** Fraud detection is a critical application in the financial sector. FL allows financial institutions to enhance fraud detection algorithms by training models on transaction data from multiple banks. This collaborative approach helps in identifying patterns and anomalies that might be missed when using data from a single source. The use of techniques like MPC and HE ensures that the transaction data remains private and secure.

**Credit Scoring:** Credit scoring models benefit from diverse data sources to improve accuracy and fairness. FL allows financial institutions to share insights without compromising privacy. Banks and financial institutions train local models on their credit data and share the updates with a central server. The aggregated model benefits from a broader dataset, leading to more accurate credit scoring. Techniques like DP ensure that individual credit data points are obfuscated, maintaining data privacy while improving model accuracy.

**Anti-Money Laundering (AML):** AML requires analyzing vast amounts of transaction data to identify suspicious activities. FL facilitates collaboration among financial institutions to enhance AML models. Financial institutions train local AML models on their transaction data and share encrypted updates for aggregation. The global model benefits from diverse data sources, improving its ability to detect money-laundering activities. Techniques like SAP and HE ensure that the transaction data remains confidential and secure throughout the process.

### 5.2.2  Ensuring Data Privacy

**Secure Aggregation:** Techniques like SAP ensure that individual financial institutions' data contributions remain confidential while still contributing to the global model.

**Differential Privacy:** Adding noise to the updates ensures that sensitive financial transactions cannot be traced back to individual users.

### 5.2.3  Case Study: Bank Fraud Detection

Fraud detection is a critical application within the financial sector, where identifying and preventing fraudulent transactions can save institutions and customers significant amounts of money and reduce the risk of financial crimes. FL provides an innovative approach to enhancing fraud detection systems by enabling banks to collaborate without exposing sensitive transaction data. This case study explores how a consortium of banks can leverage FL for fraud detection while ensuring data privacy and security.

**Background:** Fraud detection involves monitoring transactions for unusual patterns that may indicate fraudulent activity, such as identity theft, unauthorized transactions, or money laundering. Traditionally, banks develop fraud detection models based on their internal data, which limits the models' effectiveness due to the lack of diverse data sources. By using FL, banks can collaboratively train more robust and accurate fraud detection models on a broader dataset.

**Consortium Formation:** A group of banks forms a consortium to collaboratively improve their fraud detection models. The consortium establishes a FL framework that allows them to train a global model without sharing raw transaction data. The banks forming the consortium get involved in the following activities : (i) local model training, (ii) secure model update sharing, (iii) centralized aggregation, and (iv) global model distribution. These activities are discussed briefly in the following.

**Local Model Training:** Each bank trains a local fraud detection model on its internal transaction data. This process involves data preparation and model training. In the data preparation step, transaction data are preprocessed to extract relevant features such as transaction amount, frequency, location, and time of day. The model training step involves the use of ML algorithms to train the fraud detection model on the prepared data. The model learns to identify patterns indicative of fraudulent activities.

**Secure Model Update Sharing:** Once the local models are trained, each bank computes the updates to the model parameters. These updates reflect the learned patterns and insights from the local data. To ensure privacy, the updates are encrypted using SMPC and HE techniques. While Secure Multiparty Computation (SMPC) encrypts the updates so that they can be securely combined with updates from other banks, HE allows computations on encrypted data, ensuring that the updates remain confidential during aggregation.

**Centralized Aggregation:** The encrypted model updates are sent to a central server, which aggregates the updates without decrypting them. The aggregation process combines the insights from all participating banks to create a global model. Techniques like SAP and DP ensure that the server can aggregate the updates without accessing individual updates so that data privacy is protected.

**Global Model Distribution:** The aggregated global model is distributed back to the participating banks. Each bank integrates the global model with its local system, improving its fraud detection capabilities with insights gained from the broader dataset. Several data privacy and security measures are taken at this stage. Data encryption techniques are used so that all model updates are encrypted before being shared, ensuring that sensitive transaction data is never exposed. DP is used to add noise to the updates, making it difficult to trace back any information to an individual transaction. The use of SAPs ensures that the central server can aggregate the model updates without accessing individual updates, protecting the privacy of the data.

**Challenges:** The use of FL brings is several benefits in financial fraud detection in banks such as (i) improved fraud detection accuracy, (ii) enhance data privacy and security, (iii)compliance with regulations, and (iv) higher resource efficiency. However, it involves several challenges as well. Some of the challenges are (i) high technical complexity, (ii) complexity in coordination among banks, and (iii) performance and scalability issues.

*Technical complexity*: Implementing FL involves complex cryptographic techniques and secure communication protocols. Banks need to invest in the necessary infrastructure and expertise to deploy these solutions effectively. Collaboration with technology providers and research institutions can help banks implement FL frameworks. Open-source FL platforms and libraries can also facilitate the adoption process.

*Coordination complexity*: Coordinating model training and update sharing among multiple banks requires effective communication and collaboration. Ensuring that all participants adhere to the same protocols and timelines can be challenging.

Establishing a governance framework and clear communication channels can streamline coordination. Regular meetings and updates can ensure that all participants are aligned and progress is tracked effectively.

*Performance issues*: FL can introduce latency and computational overhead due to encryption and secure aggregation processes. Ensuring that the system scales efficiently with the number of participating banks is crucial. Optimizing encryption techniques and aggregation protocols can reduce latency and improve performance. Distributed computing and parallel processing can also enhance scalability.

### 5.3 Mobile and Edge Devices

FL is particularly well-suited for mobile and edge devices, enabling the training of ML models directly on devices like smartphones and IoT devices, thereby enhancing user experience while preserving privacy.

#### 5.3.1 Enhancing User Experience on Mobile Devices

**Predicting Text Input:** One of the most prominent applications of FL is in improving predictive text input on mobile devices. By training language models locally on user devices, FL allows for more personalized and accurate text predictions and autocorrect features.

**Personalized Recommendations:** FL can be used to train recommendation systems for apps, music videos, and other contents on mobile devices without sending user data to the cloud. This enhances user privacy while providing personalized experiences.

**Health Monitoring:** Wearable devices and health apps can use federated earning to improve models for health monitoring, such as detecting irregular heartbeats or predicting glucose levels, by leveraging data directly from users' devices.

#### 5.3.2 Case Study: Google's Gboard

Google's Gboard, the virtual keyboard app, is a prominent real-world example of FL in action. It demonstrates how FL can be used to improve ML models while maintaining user privacy. This case study elaborates on the implementation, benefits, and privacy measures of FL in the development of Gboard.

Background: Gboard is a widely used keyboard app that includes features like predictive text, autocorrection, and personalized suggestions. These features rely on ML models trained on user typing data to improve accuracy and user experience. However, collecting, and centralizing user data for model training poses significant privacy concerns. FL offers a solution by enabling the training of models directly on users' devices.

Implementation of FL in Gboard involves the following tasks (i) local model training on devices, (ii) model update transmission, (iii) aggregation and global model improvement, (iv) integrating privacy and security protocols and algorithms. These tasks are briefly discussed in the following.

**Local Model Training on Devices:** Instead of sending user data to a central server, Gboard trains ML models directly on users' devices. This approach involves two steps, *data Collection* and *model training*. In the data collection phase, user interactions, such as typing patterns, text inputs, and corrections, are collected. These data never leave the user's device. The Gboard app includes a local model that learns from the user's typing data. The training process occurs in the background, utilizing the device's computational resources.

**Model Update Transmission:** Once the local model is trained on the device, the updates (i.e., changes in model parameters) are sent to Google's servers. To ensure privacy, these updates are processed securely. The model updates are encrypted before transmission to protect them from interception. Only relevant and necessary updates are transmitted, reducing the amount of data sent and further protecting privacy.

**Aggregation and Global Model Improvement:** The encrypted updates from many devices are aggregated on Google's servers to improve the global model. An SAP ensures that the server aggregates the model updates without being able to view individual updates. Techniques like DP are used to add noise to the updates, ensuring that individual users' data cannot be reverse-engineered. The improved global model, which now incorporates insights from many users, is distributed back to users' devices. This model update enhances the Gboard app's overall performance and accuracy.

**Integration of Privacy and Security Protocols:** The updates from millions of devices are averaged to improve the global model. This model ensures that the data remain on the device and only model updates are shared. Standard encryption protocols like TLS (Transport Layer Security) are used to secure data in transit. Secure Multiparty Computation (SMPC): techniques are also applied to further secure the aggregation process. DP techniques are employed to add noise to the model updates. This ensures that individual contributions are obfuscated and cannot be traced back to specific users. Federated Averaging (FedAvg) is the primary algorithm used for aggregating model updates. The updates from multiple devices are averaged to form the new global model. Since only the model updates, not the raw data, are shared, privacy is preserved. Moreover, FedAvg ensures that the aggregation process is computationally efficient, allowing the system to scale across millions of devices.

**Impact:** The FL approach has significantly improved the performance of Gboard's predictive text input and autocorrect features, providing a more personalized user experience while maintaining high privacy standards. However, there are some associated challenges too. These challenges are (i) higher technical complexity, (ii) increased computational overhead, (iii) network latency and bandwidth issues.

*Higher technical complexity*: Implementing FL requires sophisticated algorithms and robust infrastructure to handle the encryption, transmission, and aggregation of model updates. Google has invested in developing and optimizing FL algorithms like FedAvg and SAPs to ensure efficient and secure implementation.

*Increased computational overhead*: Training models on users' devices can introduce computational overhead, potentially affecting device performance and battery life. The Gboard app is designed to perform training in the background, leveraging idle times and optimizing resource usage to minimize the impact on device performance.

*Network latency and bandwidth issues*: Transmitting model updates can incur network latency and bandwidth usage, especially with a large user base. Sparse and selective update transmission helps reduce the amount of data sent. Additionally, updates are often transmitted during periods of low network activity to minimize impact on user experience.

FL offers a revolutionary approach to ML by enabling collaborative model training across decentralized data sources while preserving privacy. Its application in healthcare, finance, and mobile and edge devices demonstrate the broad potential and versatility of this technology.

In healthcare, FL facilitates collaborative research and development, leading to improved diagnostic tools and personalized medicine while maintaining patient privacy. The finance sector benefits from enhanced fraud detection algorithms and credit scoring models that leverage data from multiple institutions without sharing sensitive information. Mobile and edge devices use FL to enhance user experience by training models locally, thereby preserving user privacy and providing personalized services.

Case studies like Google's Gboard and collaborative healthcare research projects illustrate the practical implementation and impact of FL. Google's Gboard demonstrates how FL can improve predictive text input on millions of devices while maintaining high privacy standards. Collaborative healthcare projects highlight the potential for FL to advance medical research and diagnostics through secure, decentralized data collaboration.

As FL continues to evolve, ongoing research and development in privacy-preserving techniques, secure aggregation, and efficient communication protocols will be crucial. By addressing the challenges and leveraging the advantages of FL, industries can harness the power of decentralized data to drive innovation, improve services, and protect user privacy.

## 6. Conclusion and Future Work

FL represents a significant advancement in the field of machine learning by addressing the crucial challenge of data privacy. This approach enables multiple entities to collaboratively train models without sharing their underlying data, thus enhancing privacy and security while maintaining model performance. Throughout this chapter, we explored the fundamentals of FL, its architecture, and workflow, and highlighted key privacy-preserving techniques such as differential privacy, encryption, and secure aggregation. Additionally, we examined the practical applications of FL in various sectors including healthcare, finance, mobile and edge devices, and industrial IoT.

The architecture of FL involves a central server and multiple local clients. The central server coordinates the overall training process, initializes model parameters, aggregates model updates, and manages communication with clients. Local clients retain their data, perform local training, compute model updates, and transmit these updates to the central server. This decentralized approach ensures that sensitive data remains localized, mitigating privacy risks associated with traditional centralized models.

Key privacy-preserving techniques discussed include differential privacy, which introduces noise to model updates to protect individual data points, and secure aggregation, which ensures that individual contributions remain confidential during the aggregation process. These methods provide robust privacy guarantees while allowing for effective collaborative training.

As FL continues to evolve, several areas require further research and development to address existing challenges and enhance its capabilities:

*Enhanced Privacy-Preserving Techniques*: Developing more robust privacy-preserving mechanisms such as advanced differential privacy techniques, homomorphic encryption, and secure multi-party computation to ensure stronger privacy guarantees.

*Improved Scalability*: Creating scalable algorithms and infrastructure to handle the massive scale and diversity of devices in cross-device FL. This includes optimizing communication protocols and reducing the computational burden on resource-constrained devices [43].

*Efficient Model Aggregation*: Innovating aggregation methods that can handle the heterogeneity of updates and improve the convergence rates of global models. Techniques like federated optimization and adaptive aggregation can play a significant role.

*Personalized FL*: Developing methods that allow the global model to be personalized for individual clients, improving performance in heterogeneous environments. Approaches like federated meta-learning and multi-task learning can be explored.

*Robustness and Security*: Enhancing the robustness of FL systems against adversarial attacks and ensuring the security of model updates. Techniques like adversarial training and secure aggregation protocols are critical [44].

*Regulatory Compliance*: Ensuring that FL frameworks adhere to data protection regulations across different regions. This involves continuous monitoring and updating of compliance strategies as regulations evolve.

*Interdisciplinary Collaboration*: Encouraging collaboration between researchers from different fields such as machine learning, cryptography, and data privacy to develop innovative solutions for FL.

By addressing the challenges mentioned above and fostering interdisciplinary collaboration, FL can continue to advance as a cornerstone of privacy-preserving machine learning. It has the potential to transform the way we approach machine learning in a privacy-conscious world, balancing the need for data-driven insights with the necessity to protect individual privacy.